\documentclass[journal]{vgtc}                
\ifpdf
  \pdfoutput=1\relax                   
  \usepackage{graphicx}                
  \DeclareGraphicsExtensions{.pdf,.png,.jpg,.jpeg} 
\else
  \usepackage{graphicx}                
  \DeclareGraphicsExtensions{.eps}     
\fi%

\graphicspath{{figures/}{pictures/}{images/}{./}} 

\usepackage{microtype}                 
\PassOptionsToPackage{warn}{textcomp}  
\usepackage{textcomp}                  
\usepackage{mathptmx}                  
\usepackage{times}                     
\usepackage{cite}                      
\usepackage{tabu}                      
\usepackage{booktabs}                  
\usepackage{enumitem}
\usepackage{dblfloatfix}
\usepackage[normalem]{ulem}

\onlineid{1175}

\vgtccategory{Research}
\vgtcpapertype{application}

\usepackage[dvipsnames]{xcolor}

\newcommand{\smolbox}[1]{{sMolBox}}
\newcommand{\manySmolboxes}[1]{{sMolBox Nodes}}
\newcommand{\smolboxes}[1]{{sMolBoxes}}
\newcommand{\inputSmolbox}[1]{{Input sMolBox}}
\newcommand{\functionSmolbox}[1]{{Measure sMolBox}}
\newcommand{\aggregationSmolbox}[1]{{Aggregation sMolBox}}
\newcommand{\unaryFunction}[1]{{Modifier}}
\newcommand{\bigBox}[1]{{bigBox}}
\newcommand{\threedview}[1]{{3D View}}
\newcommand{\importanceFunction}[1]{{importance function}}

\title{sMolBoxes: Dataflow Model for Molecular Dynamics Exploration}

\author{Pavol Ulbrich, Manuela Waldner, Katar\'{i}na Furmanov\'{a}, S\'ergio M. Marques,\\ David Bedn\'{a}\v{r}, Barbora Kozl\'{i}kov\'{a}, and Jan By\v{s}ka}
\authorfooter{
\item
 Pavol Ulbrich, Katar\'{i}na Furmanov\'{a}, and Barbora Kozl\'{i}kov\'{a} are with Visitlab, Faculty of Informatics, Masaryk University, Czech Republic. 
 \item
 Jan By\v{s}ka is with Visitlab, Faculty of Informatics, Masaryk University, Czech Republic, and University of Bergen, Bergen, Norway. \\ E-mail: byska@mail.muni.cz.
\item
 Manuela Waldner is with TU Wien, Vienna, Austria.
\item
 S\'ergio M. Marques and David Bedn\'{a}\v{r} are with Loschmidt Laboratories, Department of Experimental Biology and RECETOX, Faculty of Science, Masaryk University, Brno, Czech Republic, and International Clinical Research Center, St.\hspace{0.1em}Anne's\hspace{0.2em}University\hspace{0.2em}Hospital\hspace{0.2em}Brno, Brno, Czech Republic.
}

\shortauthortitle{Ulbrich \MakeLowercase{\textit{et al.}}: sMolBoxes: Dataflow Model for Molecular Dynamics Exploration}

\abstract{

We present \smolboxes{}, a dataflow representation for the exploration and analysis of long molecular dynamics (MD) simulations. 
When MD simulations reach millions of snapshots, a frame-by-frame observation is not feasible anymore.
Thus, biochemists rely to a large extent only on quantitative analysis of geometric and physico-chemical properties.
However, the usage of abstract methods to study inherently spatial data hinders the exploration and poses a considerable workload.
\smolboxes{} link quantitative analysis of a user-defined set of properties with interactive 3D visualizations.
They enable visual explanations of molecular behaviors, which lead to an efficient discovery of biochemically significant parts of the MD simulation.
\smolboxes{} follow a node-based model for flexible definition, combination, and immediate evaluation of properties to be investigated.
Progressive analytics enable fluid switching between multiple properties, which facilitates hypothesis generation.
Each \smolbox{} provides quick insight to an observed property or function, available in more detail in the \bigBox{} View.
The case studies illustrate that even with relatively few \smolboxes{}, it is possible to express complex analytical tasks, and their use in exploratory analysis is perceived as more efficient than traditional scripting-based methods.

} 


\CCScatlist{ 
 \CCScat{K.6.1}{Human-centered computing}{Visualization}%
{Visualization application domains}{Scientific visualization}
}

\teaser{
  \centering
  \includegraphics[width=\linewidth]{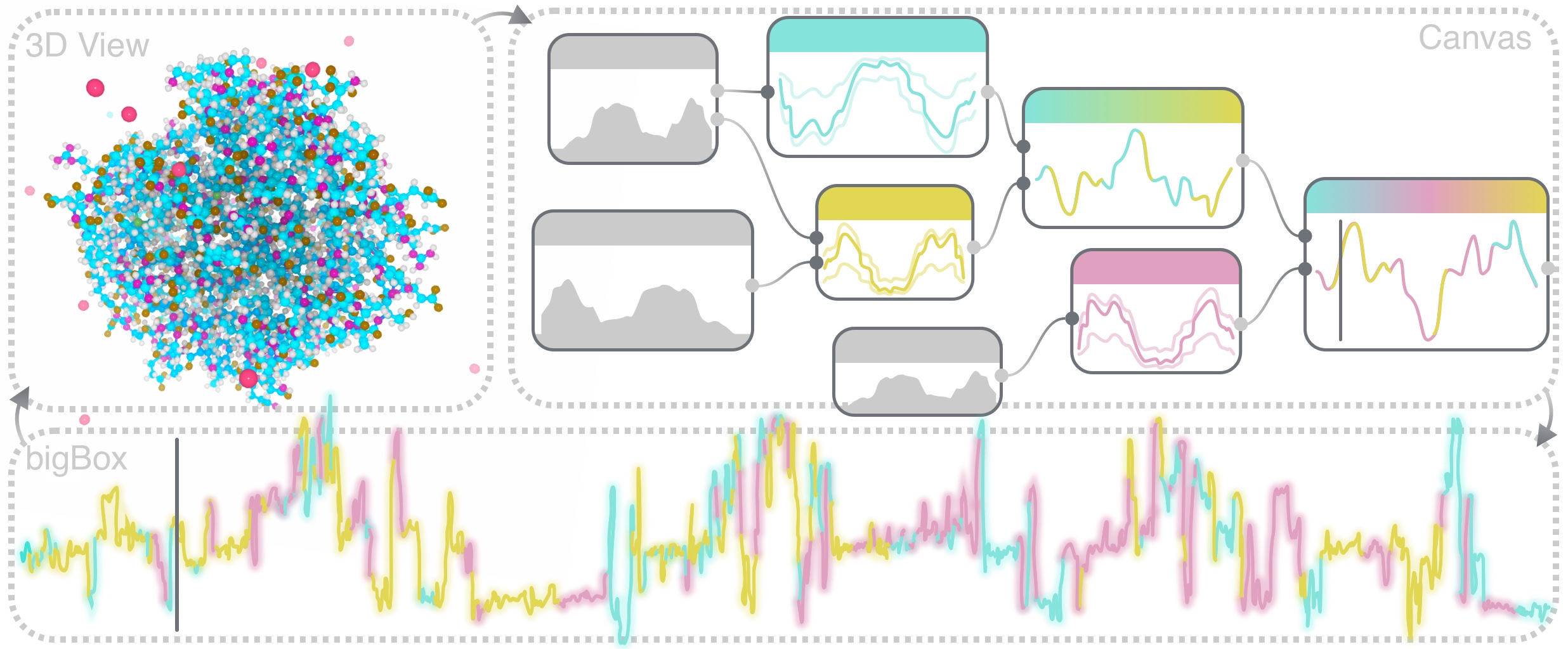}
  \caption{An overview of the \smolboxes{} interface for molecular dynamics exploration, consisting of the 3D View, the Canvas with \manySmolboxes{}, and the detailed \bigBox{} View for navigation through the 3D animation.}
  \label{fig:teaser}
}

\vgtcinsertpkg

\begin{document}


\firstsection{Introduction}

\maketitle

Studying molecular dynamics (MD) is one of the fundamental tasks in biochemical research, biotechnology, molecular medicine, or drug design.
It helps scientists understand molecules' behavior and thus improve or modify their properties.
In our research, we have been cooperating with a group of protein engineering experts for more than a decade, trying to support their workflows with dedicated visual representations. 
Their main interest lies in designing mutations of protein structures in order to change protein properties, such as its stability~\cite{koudelakova2013} or activity towards other molecules~\cite{pavlova2009}.
In order to reveal candidate amino acids for mutation and evolution, biochemists need to explore several geometric and physico-chemical properties of the protein and their changes over time.
For this purpose, molecular dynamics simulations are of the most powerful computational methods capturing molecular behavior.

Nowadays, molecular dynamics simulations span up to hundreds of thousands of time frames~\cite{Hollingsworth2018}, as they try to capture important structural changes happening in the order of nanoseconds, microseconds, or even longer time span.
The vast amount of snapshots prevents the experts from exploring and visually analyzing the simulations by traditional 3D animation techniques, commonly used for this purpose.
For example, assuming the playback speed of 5 frames per second, users would have to spend approximately 22 hours observing the animation of a simulation consisting of 400,000 time frames.
This is not feasible, and thus the exploration process has to be supported by other approaches. 
Instead of observing the whole simulation, protein engineers are trying to reveal interesting parts of the simulation, denoted as \textit{events}. 
These can be derived from various descriptive measures or features, capturing geometric (e.g., the distance between atoms, angles, root-mean-square deviation) or physico-chemical properties (e.g., electrostatic or van der Waals interaction energies, hydration) of the simulation.
These measures can be visualized in simple 2D plots where the experts are trying to reveal outliers and patterns and, even more importantly, how these measures interplay or correlate.
However, with MD simulations capturing proteins consisting of hundreds of amino acids and their interactions with multiple molecules, there are countless measures to explore. Extensive scripting would be required to generate plots for selected measures and to obtain the 3D animation sequences associated with observed patterns or events in these measures.
Consequently, protein engineers usually restrict their analysis to a few structures, such as the amino acids responsible for a reaction, making it unlikely to discover unexpected interactions or events. 

In our previously designed approach~\cite{byvska2019analysis}, we were trying to encourage exploration of the measures by providing the experts with a user interface where they could define aggregated measure functions representing the relevance of each time frame. 
Although the interface, showing superimposed functions, was well appreciated by the experts, we spotted the main drawbacks of this approach during evaluation.
The interface offered only limited features for defining aggregated functions, which proved insufficient for describing some of the more complex events in simulations. Moreover, even with the simple aggregations, it was difficult for the experts to keep track of the combinations and the provenance of functions superimposed in a single view---they quickly lost the mental model of the exploration process. 
Considering the number of values and the multi-variate and time-series character of the measures the domain experts must scrutinize, our experience revealed that appropriate visual support for this process is crucial but currently lacking.

Therefore, our work aims to fill in this gap and facilitate the complex domain-specific problem of analysis of long MD simulations. We propose a visual analysis approach supporting the interactive exploration of multiple properties and their combinations.
In order to simplify the discovery of unexpected \textit{events}, we aim to enable analysts to easily query and navigate long simulation sequences through flexible selections of spatial structures, quick computation of measures associated with these structures, and linking between 2D plots of the measures and 3D visualizations.
We call our novel solution \smolboxes{}.

We based \smolboxes{} on a dataflow network model that was already demonstrated to be intuitively understood by domain experts~\cite{waser2011nodes, inviwo, mevislab}. 
In the network, each node represents either an input, a measure, or combination of measures. By connecting nodes, users can create complex functions describing molecular behavior. Moreover, the network layout provides an overview of all functions, their settings, and aggregations at the same time.

The evaluation of \smolboxes{} was performed with a protein engineering expert, who tested the proposed interface on two exemplary scenarios from his daily workflow (presented in Section~\ref{sec:case_study}). We demonstrate that with our tool, the domain expert was able to perform the analysis in a fast manner, and that the design of the system encouraged him to explore the data beyond his typical workflow steps.
The main contributions of our work are thus the following:
\begin{itemize}
    \item Easily adaptable definitions of complex aggregated MD simulation measures that can significantly speed up previous scripting-based approaches to MD simulation analysis and support hypothesis generation.
    \item A flexible network layout providing the users with a constant overview of the provenance of the molecular interaction analysis and thus reducing their mental workload when working with complex aggregations.
    \item Progressive analytics of large MD trajectories facilitating immediate previews of calculated measures and functions refined in an iterative manner.
    \item Direct linking of the constructed molecular interaction analysis network with a 3D representation of the protein, enabling the selection of inputs and qualitative inspection of outputs, which is unavailable when using standard statistical analysis.
\end{itemize}

\section{Data---Tasks---Requirements}
\label{Sec:Requirements}
In this section, we first describe the \textit{data} we are working with and the related domain-specific terminology.
Then we discuss the workflow traditionally followed by the experts when analyzing MD simulations and the main \textit{tasks}, given by the domain experts, that we are aiming to support with this work.
Finally, we discuss the design \textit{requirements} for our system that we derived from the tasks.

\begin{figure}[tb]
  \centering
  \includegraphics[width=0.9\linewidth]{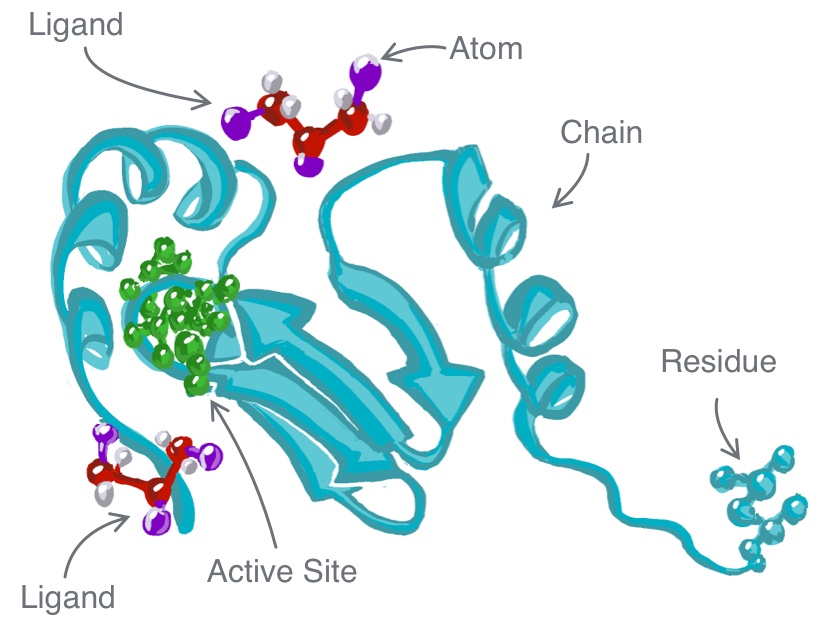}
  \caption{Illustration of input molecular structures associated with molecular dynamics simulations.}
    \label{fig:data_overview}       
\end{figure}

\subsection{Data and Terminology}
As already stated, we are focusing on the analysis of molecular systems with thousands of atoms and molecular dynamics simulations of their movements, which can consist of hundreds of thousands of time frames. Due to the amount of data, binary formats which store only information about the position of atoms over time are often used. Therefore, some properties, like mutual distance or energy between atoms, have to be extracted afterward. For some of these properties, this may require nontrivial computation time.
While our research is primarily driven by the needs of biochemists, the proposed solution can be applied to similar datasets consisting of positional information of particles over time. 

Molecules can be further described as multi-scale structures where individual atoms form \textit{amino acid residues} (often referred to as \textit{residues}), which are then connected into linear structures called \textit{chains} (see Figure~\ref{fig:data_overview}).
Each molecule in the simulation then consists of one or multiple of these chains.
Additionally, the simulation may contain water molecules forming the \textit{solvent} (outer environment of the protein) or other small molecules called \textit{ligands}.
These are playing important role in protein reactivity.
In protein-ligand simulations, ligands bind to the \textit{active site} of the protein (see Figure~\ref{fig:data_overview}), and protein engineers analyze which characteristics and properties influence the success and speed of this binding.
The simulation of binding can be augmented with additional time-series data computed with different tools, such as binding energies or various properties computed for the hollow cavities (e.g., \textit{tunnels})~\cite{brezovsky2018computational} inside the protein, often serving as access pathways for the ligand to the active site. The narrowest part of the tunnel is called the \emph{tunnel bottleneck} and it is formed by multiple \emph{bottleneck residues}.
The radius of the tunnel bottleneck varies over time and its opening and closing has impact on the tunnel's permeability. 

\subsection{Tasks}
Our proposed work directly builds on lessons learned from our previous work~\cite{byvska2019analysis}, where we have identified several requirements for the analysis of long MD simulations in protein engineering based on 3D animations.
However, the aforementioned work primarily addresses design considerations for spatio-temporal focus+context in the 3D animations and offers very limited flexibility for expressing the investigation focus.
Our previous requirements can therefore be considered orthogonal to the requirements of the present work.

Within the period of almost two years, we conducted numerous sessions with always at least six protein engineering experts on different levels of seniority (two senior researchers, two postdocs, and Ph.D. students).
Of these experts, one senior researcher and one postdoc are co-authors of this paper.
These sessions had two main goals: 1)~shadowing, i.e., observing the experts in their daily routine with the existing workflows and tools; 2) discussing the benefits and limitations of the workflows and tools, which could be addressed by visualization.

Within these sessions, we specifically analyzed three works published by the protein engineering experts, which focused on different aspects of molecular behaviour~\cite{marques2017catalytic,kokkonen2018molecular,marques2017enzyme}.
Based on the analysis of these cases and our observations, we have identified the following tasks for our proposed solution:

\begin{enumerate}[label=T\arabic*:,leftmargin=0.65cm, parsep=1pt]

\item \textbf{Property exploration.} The biochemists typically start the analysis of MD simulations by exploring various properties that could navigate them to important snapshots of the simulation, such as distances or angles between residues~\cite{byvska2019analysis}.
Another important aspect is also the correlation between these properties.
One example of such a case is the identification of gating residues, which control the access to the protein's active site~\cite{kokkonen2018molecular}.
Gating residues usually exhibit complex movement, often in concordance with other residues.
In this case, the biochemists would explore the root-mean-square deviation (RMSD) of the entire protein or movable parts of the protein, the distance between various residues, their angles, and the correlation between these measures on multiple residues and substructures of the protein.
In their traditional workflow, experts calculate the bulk or the time-evolution of the properties using in-house bash or python scripts, jupyter notebooks, or Markov state models in order to cluster the conformational states and estimate their properties. 
All these analyses are hypothesis-driven, and, due to the large amount of data that is produced, it is easy to overlook relevant details not considered beforehand.
The number of properties and residues the biochemist usually explore is limited to a certain extent.

\item \textbf{Identification of important snapshots.} Once the domain experts identify the most important properties, they want to use them to identify the relevant snapshots of the simulation, i.e., time-stamps carrying potential causes for the observed values or change. At this stage, they typically formulate a hypothesis about their observations.
For example, RMSD changes indicate conformation changes of the protein, and through visual inspection, it may be possible to speculate what might have triggered them. Another example is finding a geometric arrangement of the protein in which a reaction could take place~\cite{marques2017catalytic}.
Such molecular conformation can be identified as a snapshot, in which the distance between two selected molecular structures is below a certain threshold, and the angle at which they approach is within an acceptable range.

\item \textbf{Causality and conditions analysis.} Finally, when the important snapshots are identified, the biochemists study the conditions and causes of the given state in greater detail to validate their hypothesis. 
Here, they sometimes extend their analysis with new properties and structures, if they suspect these can play a role in the inspected event.
Additionally, the retrospective analysis of the properties can help them identify possible modifications to the protein that could lead to change in its behavior.
An example could be a discovery of residue that is preventing the binding of a ligand~\cite{kokkonen2018molecular}.
Such residue could then be experimentally replaced to increase the reactivity of the protein and enable the entrance of a ligand into the active site.

\end{enumerate}

These tasks are typically performed iteratively. Through visual analysis of causality and events (T3), interesting structures and behaviors might be discovered and subsequently analyzed in more detail for their properties (T1).
However, since tasks T1 and T2 are tedious to perform with traditional tools, T3 is done relatively rarely nowadays. 

\subsection{Requirements}

Based on the stated tasks and the limitations we identified in the current workflows of protein engineers, we derived the following design requirements for our system: 

\begin{enumerate}[label=R\arabic*:,leftmargin=0.65cm, parsep=1pt]

\item \textbf{Selection of spatial inputs on multiple levels of detail}, e.g., atoms, residues, or even entire molecules. To start the analysis~(T1), the system should allow the users to flexibly select, modify, and combine molecular (sub)structures to be analyzed.

\item \textbf{Predefined measure functions} that compute a selected metric on one or multiple spatial inputs for each snapshot without the need for scripting. This is important for initial exploration of the properties (T1), but the measure functions also serve as indicators of snapshot importance (T2). The system should support a range of measures typically associated with MD simulation analysis, such as the distance between two spatial inputs or RMSD.

\item \textbf{Flexible aggregations} and combinations of the measure functions. The system should support the aggregation of multiple measures with aggregation operations, such as extrema or average, to combine the measure functions into a single output defining the investigation focus (T2).

\item \textbf{Data provenance representation} to support analytical scenarios where the exploration might lead to complex aggregated functions that are composed of multiple inputs and many different measures. In such cases, it is impossible for the users to keep the mental model of the constitution of aggregated functions.
Nevertheless, it is vital for the analysis of the causes and effects (T3) of MD simulation phenomena. The system should provide a representation of the function model and the effects individual components have on the resulting function.

\item \textbf{Real-time performance} of the structure loading and function computation. 
These operations need to be performed without any significant latency to enable a fluid analysis of the desired molecular properties (T1, T2).

\item \textbf{Visualisation of large time-series data} containing hundreds of thousands of snapshots. Due to their size, measures extracted from long MD simulations cannot always be displayed in their full resolution and require dedicated solutions for their visualization~(T1-T3). 

\item \textbf{Linking of measures with the 3D representation} to support tight coupling between qualitative inspection and in-place analysis of interesting observations and in-depth quantitative analysis, since both are needed for investigation of causality in MD simulations~(T3).

\end{enumerate}

\section{Related Work}
In this work, we present a system for the visual analysis and exploration of MD simulations. Therefore, this section will first discuss related visual analytics approaches from this domain area.
Furthermore, in our presented work, we take inspiration from dataflow models, which have already been successfully used in several applications for analytics in various domains, including life sciences. Therefore, in the second part of this section, we discuss these approaches.

\subsection{Visual Analysis of MD Simulations}
An extensive study performed by Hensen et al.~\cite{Hensen2012} proved that there is a strong correlation between the dynamics of molecular structure and its function. Therefore, without a proper understanding of protein dynamic behavior, we cannot fully understand its function.
There are already many existing approaches aiming to provide insight into MD simulations~\cite{martinez2020visualizing}.

One possible approach is to aggregate the desired information about the simulation and represent it in a static overview visualization.
Patro et al.~\cite{patro2010} measure the saliency in MD simulations which guides the selection of representative
and anomalous time frames summarizing the MDs.
In their following research~\cite{patro2011}, they represent the MD simulation as a state transition graph, illustrating the flow of a biomolecule through the trajectory space.
Bryden et al.~\cite{bryden2012} use arrow glyphs to depict the molecular flexibility calculated from normal mode analysis. 
First, they cluster atoms according to their common movement patterns. These clusters are then annotated with arc-shaped arrows in 3D representation of the molecule.
When the simulation contains many spatially-distributed trajectories of molecules, such as lipids, the overview representation often utilizes streamline-based representations, such as in MolPathFinder~\cite{Alharbi2016} or in the approach proposed by Ertl et al.~\cite{Ertl2014}.
In the latter, the authors perform a spatial and temporal aggregation of ions transported through a porous structure, resulting in a velocity field visualization.
Bidmon et al.~\cite{bidmon2008} present an approach to visual abstraction of solvent paths traversing through a protein. 
Their aggregated visualization also encodes information about the direction and velocity of the solvent.
The overview is reached by clustering similar pathlines and showing only principal paths.
Vad et al.~\cite{vad2017watergate} also focus on studying solvent paths when they propose a set of visualizations for interactive filtering and exploration of solvent trajectories and observing the space inside protein occupied by solvent over time.
Schatz et al.~\cite{schatz2021visual} aggregate information about molecular interactions in the whole MD simulation and display it in an enhanced amino-acid sequence diagram. However, in this case, the temporal aspect of the simulation is lost.
In general, the overview visual representations of the MD simulations are very powerful and informative, giving the user the first insight without tedious frame-by-frame exploration.
On the other hand, to the best of our knowledge, none of the existing approaches is suitable to facilitate a multifaceted interactive exploration of the MD simulation using traditional visual encodings for geometric and physico-chemical measurements.

There are also many existing approaches with the main focus on simulations of only a few molecules or particles and a more detailed exploration of changes in their properties over time.
Furmanov\'{a} et al.~\cite{Furmanova2017} present a visual analysis tool for detailed exploration of geometric and physico-chemical properties of individual ligand paths inside a protein.
Although this approach could be useful in our case, the main limitation is the narrow focus of this tool---it can analyze only a single ligand path over time. 
Similarly, V\'{a}zquez et al.~\cite{vazquez2018} study the path of a single ligand through the protein in a very abstracted way. 
Even though the proposed visualization is very space-saving and enables plotting many properties simultaneously, it only relies on abstract visualizations. It does not support tight coupling between abstract and 3D visualization for more exploratory analysis.
Additionally, it does not intuitively allow combining these properties to detect important events in the MD simulation.
This is partially solved in the approach by Duran et al.~\cite{duran2019}, which enables to explore a limited set of properties of more ligands at the same time.
However, for each ligand, the tool creates an individual plot, and therefore, it is not extendable to more complex tasks, such as analyzing many ligands and solvents at the same time.
Sk\r{a}nberg et al.~\cite{Skanberg2018} are following a similar idea and convey selected properties in 2D plots.
Additionally, they show the density field depicting the spatial distribution of molecules within the simulation. 
However, the support for combining properties and thus deriving important events is missing.
As we observed in our previous study~\cite{byvska2019analysis}, individual measures are not sufficient to describe complex events or behaviors, which we aim to address using \smolboxes{}.

\vspace{2mm}
\subsection{Navigation Systems for Visual Exploration}
In our solution, we are building upon so-called modular visualization environments~\cite{Upson1989}, where the user interacts with the system by directly manipulating objects in the scene.
Combining these objects creates the mental map and resulting visual representation of the whole analysis process.
Our work was inspired by visualization systems relying on dataflow diagrams to facilitate interactive visual exploration.
One prominent example is the system by  Waser et al.~\cite{waser2011nodes}, who proposed a dataflow diagram for interactive exploration of the simulation setup of the flood emergency simulation system.
The system consists of a set of nodes, in which each node type can contain a set of connectors, serving to establish a relationship with other nodes.
However, the complexity of the studied tasks often results in a complicated visual interface, as the nodes can also encode quite complex information, such as a 3D View.
Dataflow diagrams were also employed by Yu and Silva~\cite{Yu2017}.
They use them for specifying relations between components that process, filter, or visualize the data and can be combined by interacting with the visual interface.
Similar examples are \emph{DataMeadow}~\cite{elmqvist2008datameadow}, \emph{FlowSense}~\cite{yu2019flowsense}, or \emph{ExPlates}~\cite{javed2013explates}.
These systems typically operate by connecting data source nodes with multiple data operators and visualization target nodes, which specify the visual mapping~\cite{mei2018design}.
In contrast, \smolboxes{} do not specify visualizations but rather serve as navigation aid through the long trajectories and for highlighting within the crowded molecular scene.
The 3D visualization itself relies on conventional visual mappings for molecular data, such as balls-and-sticks or ribbon representation.

In medical imaging, the modular visualization environment was used in the MeVisLab tool~\cite{mevislab}.
Such an environment also forms an integral part of the system presented by Meyer-Spradow et al.~\cite{spradow2009}.
Their Voreen framework consists of so-called processors---autonomous functional building blocks serving for specific volume rendering tasks.
By combining these blocks into dataflow networks, the user can define the visual appearance of the rendered object.
A similar idea was used already before in the VisTrails~\cite{Silva2007} toolkit. 

The concept of modular environments is followed not only by researchers solving tasks using visual interfaces, but the same principle is used in visual programming.
Here, instead of writing lines of codes, users are generating the program by visually connecting nodes with specific functions.
The principles of visual programming are well implemented in several tools across multiple domains, such as VVVV~\cite{vvvv} for graphics and media, LabVIEW~\cite{kodosky2020labview} for engineering, and SimuLink~\cite{dabney2004mastering} for simulations.

Recent development can be seen in the Inviwo tool~\cite{inviwo} and its direct application into molecular exploration in the VIA-MD software~\cite{Skanberg2018}.
However, these tools do not rely on a dataflow model in true sense. Instead, they are limited to displaying individual metrics and are not designed to describe rather complex semantic behaviors.
In addition, they offer only limited linking between the 3D view and the 2D time-series plots.

In summary, to the best of our knowledge, dataflow models have not yet been used for the exploratory analysis of long MD simulations.
 
\section{\smolboxes{} Design}

\begin{figure}[b!]
  \centering
  \includegraphics[width=1\linewidth]{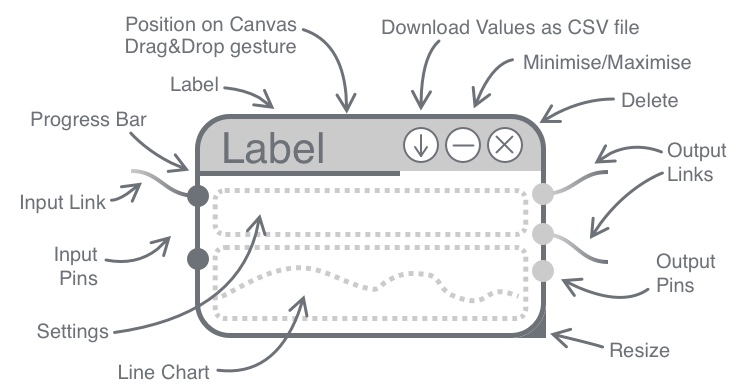}
  \caption{Illustration of the general sMolBox layout and its components.}
    \label{fig:sMolBox_structure}
\end{figure}

\begin{figure*}[t]
  \centering
  \includegraphics[width=0.95\linewidth]{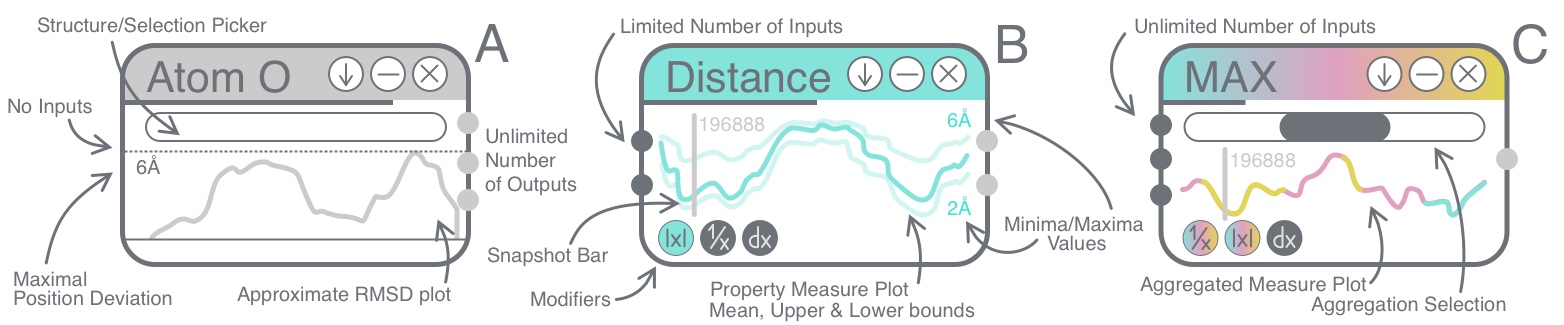}
  \caption{\label{fig:sMolBoxes}
            Each sMolBox has different layout, based on its function: (left) Input sMolBox, (middle) Measure sMolBox, and (right) Aggregation sMolBox.}
\end{figure*}

\begin{figure}[ht!]
  \centering
  \includegraphics[width=1.0\linewidth]{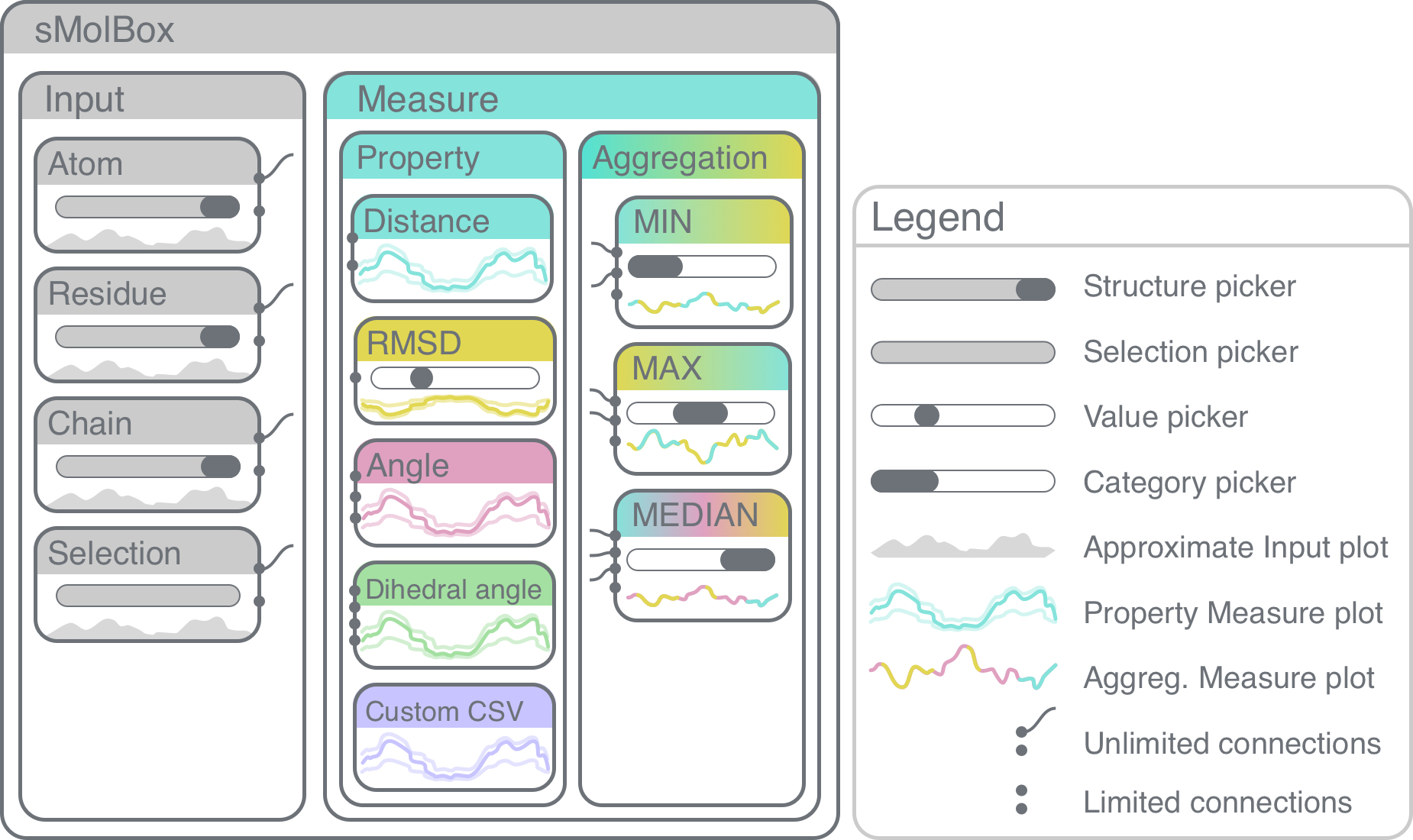}
  \caption{Overview and categorization of the sMolBox instances.}
    \label{fig:categorization}  
\end{figure}

Our tool is conceptually similar to dataflow networks. We represent each node in the network by a \textit{box} that we call \smolbox{} (see Figure~\ref{fig:sMolBox_structure}). 
By analyzing the workflows of our collaborators (see Section~\ref{Sec:Requirements}), we concluded that we could support their complex analyses with just three types of nodes.
The \smolboxes{} act as input selectors, property measures, or aggregation functions.
By forming connections between these boxes, users can easily and quickly explore various biochemical properties and their combinations (T1).
The network also produces an \textit{\importanceFunction{}} expressing the user-defined notion of relevant events over time. This function can be used to guide the analysis and navigate the users to the most interesting parts of the data (T2). Finally, since the network can be easily adjusted, additional measures can be added to study the possible causes of the identified events and observations in greater detail (T3).

The three types of nodes (Figure~\ref{fig:sMolBoxes}) in the \smolbox{} network have the following roles.
We need \textit{\inputSmolbox{}} for specifying the spatial data to be analysed (\textbf{R1}), \textit{\functionSmolbox{}} for calculating various properties of the selected data (\textbf{R2}), and \textit{\aggregationSmolbox{}} for combining the properties (\textbf{R3}).
The network layout of \smolboxes{}, enhanced by visual cues, naturally provides an overview of data provenance (\textbf{R4}).

To provide users with a consistent visual design, all \smolboxes{} have a similar layout with a preview line chart as the main element, as shown in Figure~\ref{fig:sMolBox_structure}.
To motivate the users to build the dataflow network from left to right so they utilize the available space most efficiently, we place the connection pins on the sides---input ones on the left and outputs on the right. 
The boxes may be linked by an intuitive drag\&drop interaction with the pins.
The system prevents faulty connections that would violate the correct flow of data, such as an \inputSmolbox{} following a \functionSmolbox{}.
In order to highlight the flow of data in complex networks, we color the links with a lighter color near the parent box and a darker color near the child box.
All \smolboxes{} are designed to be fully responsive, and all their components, including the time-series visualizations, are computed in real-time, in parallel, and are not blocking any other process (\textbf{R5}). We discuss the progressive data processing approach that facilitates this in Section~\ref{Sec:real-time-analytics}. 

The final application consists of three connected views (see Figure~\ref{fig:teaser}): Canvas, \bigBox{} View, and 3D View. The \textit{Canvas} is the analyst's central workbench for adding, connecting, and managing individual \smolboxes{}.
When the analyst selects a \smolbox{} on the Canvas, the \textit{\bigBox{}} View shows the corresponding measure in detail. Finally, the \bigBox{} or any \smolbox{} can be used to control the linked \textit{3D View} for qualitative inspection of behaviors observed in the measures (\textbf{R7}).
Below, we describe the three types of \smolboxes{} and their interplay with the proposed views in more detail. 

\subsection{Input sMolBoxes}
Protein engineers often start their analysis by investigating the behavior of ligands or relevant parts of the protein, such as protein residues directly involved in chemical reactions (e.g., \textit{catalytic residues}).
To allow a direct and quick selection of molecules or their parts on multiple levels of detail (\textbf{R1}), we provide \inputSmolbox{}es (see Figure~\ref{fig:sMolBoxes} left). 

Using \inputSmolbox{}, the analyst can select a single atom, residue, or molecular chain by typing its respective ID or selecting it directly in the 3D View. 
To support the exploration of larger areas of interest, e.g., multiple catalytic residues at the same time, we also have to provide users with the means to create arbitrary spatial selections anytime during their exploration.
Here we benefit from the existing functionality inside the CAVER Analyst software~\cite{jurcik2018caver}, where we implemented our prototype.
Users can perform complex selections through direct interaction in the 3D View, by interacting with the Sequence View showing the color-coded list of amino acids forming the protein, or by typing commands in the built-in console.
The selections can be then mapped to \inputSmolbox{}es.
To enable users to easily relate \inputSmolbox{}es to the corresponding parts of the molecule later in the analysis process (\textbf{R4}), we highlight the corresponding parts in the 3D View every time the user clicks on an \inputSmolbox{}.

Each \inputSmolbox{} provides a preview line chart indicating to the users potentially interesting time frames (\textbf{R2}) with significant conformation changes already during the data selection process. Conformation changes indicate that the structure has changed its shape or position; they may therefore represent an interesting behavior to investigate in more detail.
For \inputSmolbox{} with a selected atom, this line chart corresponds to the RMSD of the atom with respect to the first snapshot of the simulation.
For inputs consisting of multiple atoms, the RMSD is approximated (see Section~\ref{sec:approximation}) to provide a quick indication of the structure's spatial development over time. 
To allow users to estimate the size and the relevance of these conformation changes, we display upper and lower bounds in Ångström units next to the line chart.

\begin{figure*}[t]
  \centering
  \includegraphics[width=1.0\linewidth]{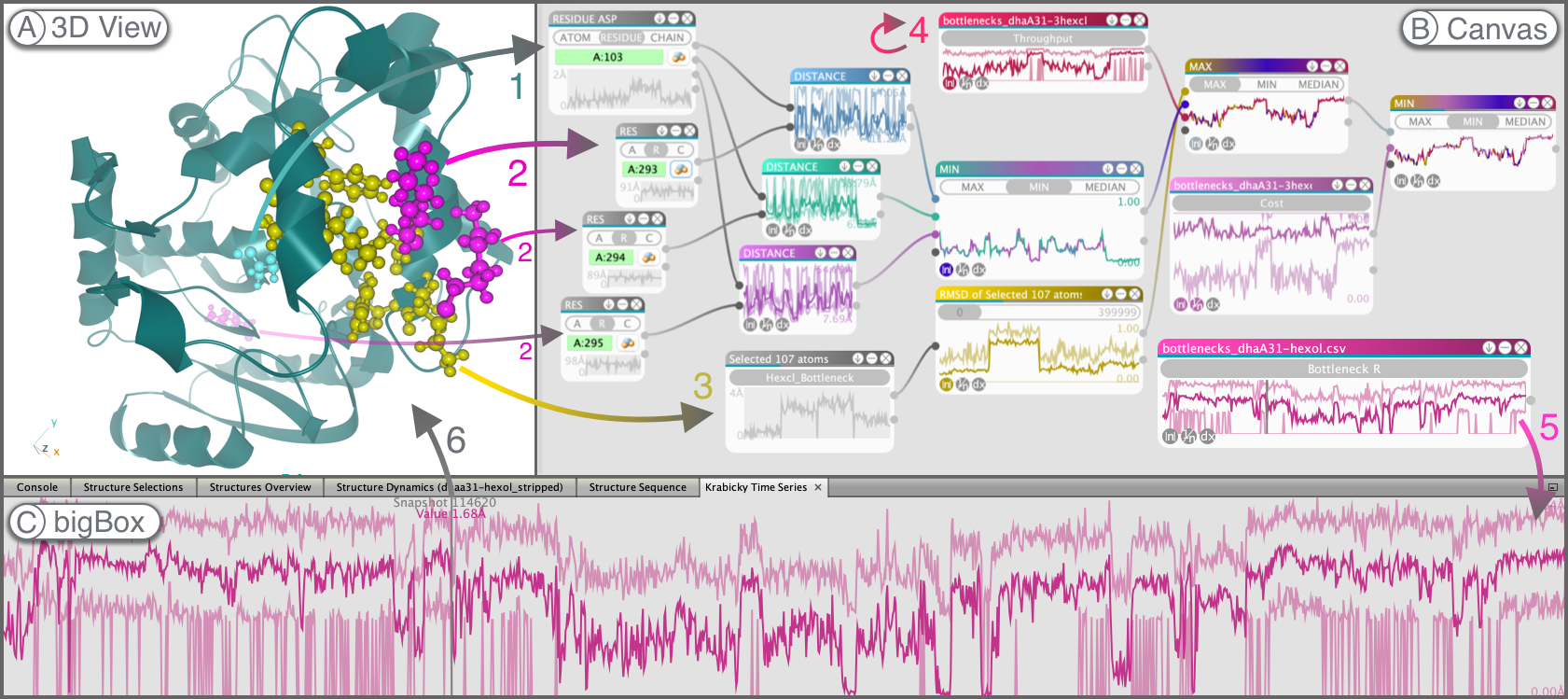}
  \caption{\label{fig:workflow}
            Analysis of the release of ligands in an MD simulation with 400,000 snapshots.
            We selected an active site residue (1) and three ligands (2) in the 3D View (A), then we calculated their minimum distance (blue, green, and purple distance sMolBox) using MIN aggregation sMolBox.
            We visually compared the proximity to the active site with the RMSD (yellow sMolBox) of the bottleneck residues (3) and the bottleneck throughput loaded as an external file (red sMolBox, 4).
            After normalization, we were able to use the MAX sMolBox to evaluate the most important time-frames in the simulation with respect to all the measures. 
            Subsequently, we utilized the MIN sMolBox to investigate the role of tunnel cost measure on the interesting snapshots.
            It led us to the part around snapshot 114,620, where we suspected a movement of ligands through the protein bottleneck.
            The detailed examination of the tunnel radius (5) using bigBox (C), followed by a more precise observation in the 3D View (A), has shown a sudden entry and exit of one of the ligands to the active site.
            This observation would not be feasible with a mere analysis of a distance graph.}
            \vspace{-2.5mm}
\end{figure*}

\subsection{\functionSmolbox{}es}
\functionSmolbox{}es compute geometric properties on the connected \inputSmolbox{}es for every snapshot. This way, the analyst can start extracting properties describing the behavior of molecules or their parts in time (\textbf{R2}).
\functionSmolbox{}es require one or multiple \inputSmolbox{}es depending on the function they are computing. For example, one \inputSmolbox{} is enough to compute \emph{RMSD}, but \emph{distance} requires two inputs.

A major part of the \functionSmolbox{} space is occupied by the line chart depicting the computed measure over time (see Figure~\ref{fig:sMolBoxes} center).
We have opted for a line chart as it is easy to understand, provides a well-known representation of trends, and is also predominantly used by protein engineers in their traditional analyses.
However, given the amount of data, it is not possible to accurately display all values, as it would require hundreds of thousands of horizontal pixels (\textbf{R6}). 
Therefore, we bin the function by the number of available pixels, and we visualize the average and the upper and lower bound of the values within each bin using polylines.

To enable qualitative inspection of behaviors observed in the line chart, we interactively linked it with the 3D View via the \textit{Snapshot~Bar}~(\textbf{R7}).
This vertical bar is situated over the line chart in the approximate position of the currently depicted snapshot in the 3D View.
The bar position, and thus the snapshot depicted in the 3D View, can be adjusted by dragging the bar along the chart. For binned functions, re-positioning of the bar selects the first snapshot from the corresponding bin.
We depict the current snapshot number next to the bar for better orientation. If the \functionSmolbox{} is big enough, we also show the calculated value of the displayed measure.

Each \functionSmolbox{} is given a unique color, which is depicted on top of the \smolbox{} and in the line chart representation. This can be seen in Figure~\ref{fig:categorization}, which provides the overview of all \smolbox{} instances supported by our tool. We decided to use color for \functionSmolbox{}es and not for \inputSmolbox{}es because the measured property gives a particular meaning to the input. For example, having two residues, we can measure their distance, binding energy, or other biochemical properties. If we colored respective \functionSmolbox{}es by their input, they would all have the same color. However, by assigning a unique color to each measure, we ensure they are easily recognizable from each other in the network. Moreover, this way, we can also better communicate the provenance of combined measures (\textbf{R4}), as will be described in the next section.

\subsection{\aggregationSmolbox{}es}

A major challenge when exploring MD simulations is the combined analysis of various measures.
For example, protein engineers may want to detect when any of the ligands present in the simulation enters the protein active site.
This process requires computing the distance between every ligand and the active site and consequent comparison of obtained measures (\textbf{R3}).
To support such analyses, we designed \aggregationSmolbox{}es.
They accept an unlimited number of \functionSmolbox{}es and \aggregationSmolbox{}es as input and calculate a new function as a combination of all parent measures according to the selected type. Based on discussion with the domain experts, we have selected the following three aggregation functions that are most often used in their everyday workflow: \textit{Minimal}, \textit{Maximal}, and \textit{Median} function.
The type of the \aggregationSmolbox{} can be changed at any time using a horizontal categorical selector (see Figure~\ref{fig:sMolBoxes} right). 

\aggregationSmolbox{}es are grayscale at first, but they inherit the colors from their parent \functionSmolbox{}es (or \aggregationSmolbox{}es). To ensure the users can keep an overview of measures forming the aggregated function (\textbf{R4}), the inherited colors are used for the top part of the \aggregationSmolbox{}, connection links, and the final aggregated function depicted as a line chart. The colored links help with better navigation in the network representation. Furthermore, by coloring the function depicted in the \aggregationSmolbox{}, we convey to the user the time intervals where, for example, in the case of maximum aggregation, a particular measure was dominating (see Figure~\ref{fig:workflow}). 

Instead of showing upper and lower bounds, the line chart in \aggregationSmolbox{} focuses solely on the average value of the aggregated bins.
We opted for this solution because the maxima and minima for the aggregated intervals would not be connected into a smooth trend line and would increase clutter in the visualization.
Similarly to the \functionSmolbox{}, also here, the users can utilize the Snapshot Bar for easy navigation through the time series.

\begin{figure*}[t]
  \centering
  \includegraphics[width=1.0\linewidth]{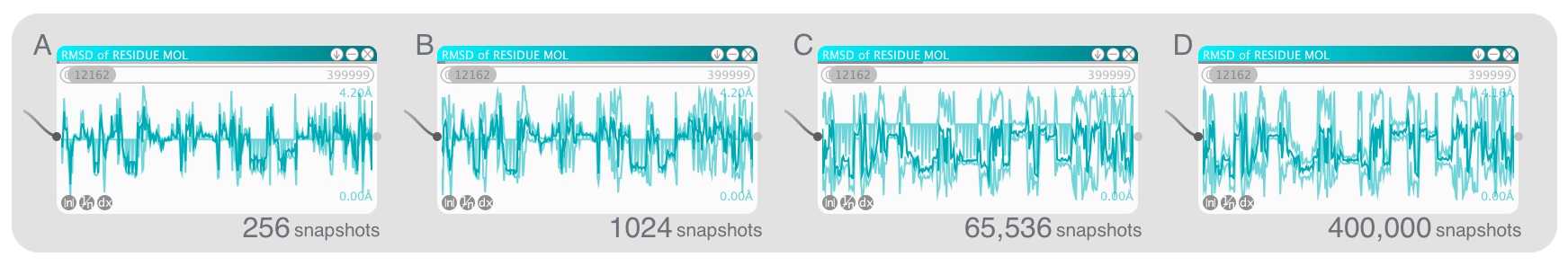}
  \caption{\label{fig:pa}The progressive calculation of RMSD for a ligand demonstrates the change in the line chart for 256 (A), 1024 (B), 65,536 (C), and 400,000 processed MD snapshots.
            }
            \label{fig:progressiveRMSD}
\end{figure*}

\subsection{\unaryFunction{}s}
For some analytical tasks, smaller values of the measured properties are more important (e.g., a small distance between protein and ligand residues implies an interaction). In other cases, larger values of the same measure are of interest (e.g., an increasing distance between the residues may imply the opening of the protein tunnel). To support these analytical tasks, both \functionSmolbox{}es and \aggregationSmolbox{}es offer a set of \unaryFunction{}s---linear unary functions that take the computed \importanceFunction{} as a single input and adjust all the values according to the \unaryFunction{} definition. We have implemented the following three \unaryFunction{}s:

\begin{itemize}[topsep=1pt,itemsep=0ex, parsep=0ex]
    \item \textbf{Normalization} can be used when the analyst is trying to compare seemingly incomparable measures, such as the distance between two atoms of the same residue, typically ranging in low numbers, and the distance between molecule and ligand that can be in the order of magnitudes higher. With normalized values, users can compare relative changes in the measures.
    \item \textbf{Inverse} can be used in case the analyst wants to compare a maximum of one property measure against a minimum of another, which can be easily flipped using this \unaryFunction{}.
    \item \textbf{Derivation} can be used to identify rapid changes in the measure represented by the \smolbox{}.
\end{itemize}

\unaryFunction{}s are represented by circular icons at the bottom left corner of \smolboxes{} (see Figure~\ref{fig:sMolBoxes} center and right). We opted for this solution because the \unaryFunction{}s will always have only one input, and having them as separate nodes would unnecessarily complicate the final network. It also provides means for the users to quickly perform fast analysis of the measures, for example, by exploring the rate of change of the corresponding measure via derivatives. 

In the beginning, all \unaryFunction{}s are in a non-active state. They can be activated by clicking on their corresponding icons.
Since not all \unaryFunction{}s are commutative, their order might also be altered by easy drag\&drop interaction.

\subsection{\bigBox{}}
By connecting individual \smolboxes{}, users can quickly explore various biochemical properties and their combinations (\textbf{R1-R3}). However, due to their restricted size, \smolboxes{} can only provide a limited view of the investigated time series, the lengths of which can reach hundreds of thousands of time steps (\textbf{R6}). To support a detailed analysis of these more extensive measures, users can select any \smolbox{} as their current analysis focus. We then depict the corresponding measure in the linked \bigBox{} View, which utilizes the whole width of a screen and thus shows the selected \importanceFunction{} in a better temporal resolution.

The \bigBox{} also serves as the primary interface to control the 3D View. It thereby replaces the classic VRC-controls to play and pause the animation and the traditional navigation bar for scrubbing through the animation without any visual guidance. The \bigBox{} supports scrubbing with respect to measure associated with the selected \smolbox{}. This way, the displayed measures can be qualitatively explained with the associated 3D visualization (\textbf{R7}).

\subsection{3D View}
As noted in our previous focus groups~\cite{byvska2019analysis}, 3D animations are useful to qualitatively characterize quantitative observations and to get a better intuition about spatial relations. For example, exploration of the RMSD for a molecule only indicates that the conformation changed, but it does not convey any information on \emph{how} it changed.
With the linked 3D View, it can be checked effortlessly (\textbf{R7}).
However, dense molecular environments pose a significant challenge because the most interesting parts of the molecule may be invisible due to occlusions.
\smolboxes{} take liberty on several standard molecular representations with various levels of abstractions (e.g., van der Walls, ball-and-stick, cartoon, or backbone representations) implemented in CAVER Analyst software~\cite{jurcik2018caver}.
The user can choose different representations for the spatial selections represented by Input sMolBoxes. This way, users can visualize the selected structures in full detail while keeping the rest of the molecule abstracted as context.

Linking the Canvas with the 3D View enables a new workflow, where analysts can fluidly switch between rapid quantitative analysis of structures and functions and more qualitative exploration of molecular behaviors triggering these functions.
The three main views---Canvas, \bigBox{}, and 3D View are connected through a set of interactions.
Upon a close examination of the nuances of a calculated function, it is suitable to explore the 3D conformations in the snapshots of interest.
Therefore, after clicking on a calculated function, the 3D View is set to display the corresponding snapshot.

\section{Real-Time Analytics}
\label{Sec:real-time-analytics}
Exploratory analysis requires real-time selection and computation of measures to enable a fluid, uninterrupted workflow (\textbf{R5}).
It poses a substantial challenge for long MD simulations consisting of tens of thousands of atoms and hundreds of thousands of snapshots. 
Exploratory analysis is, per definition, unpredictable, as even the analysts themselves do not initially know the objectives of their investigation.
Pre-computation of all possible selections and their associated measures is therefore not feasible.

Dataflow models, by design, lend themselves to real-time analytics since their scope is clearly defined, and their associated actions, such as loading and computing, can be performed in parallel.
Additionally, we introduced several measures to enable real-time interaction and analysis, including progressive loading, predictive caching, approximate computations, and selective pre-computations of very costly functions.

\subsection{Progressive Loading and Computation}

I/O operations are the most costly among all \smolbox{} types, as they have to load the positional vector data from the slow memory into the fast cache.
To minimize the latency, atom positions are loaded in an iterative progressive manner, where each iteration adds twice the amount of data of the previous iteration (see Figure~\ref{fig:progressiveRMSD}). 
Empirically, we found 256 snapshots uniformly distributed across the simulation to be a reasonable initial number of snapshots so that we can provide the first overview of the data in the \smolbox{}es instantly.
Since every calculation is being processed on a separate thread, the user can interact with the system without interruptions.
The same principle applies also to the aproximation of the \importanceFunction{} in \functionSmolbox{}es. Every new iteration is processed as soon as it is available by the parent \inputSmolbox{}.
Every \smolbox{} also contains a progress bar (see Figure~\ref{fig:sMolBox_structure}), indicating a ratio of dynamic snapshots that were already processed by it.

\subsection{Predictive Caching}
\label{sec:predictiveCaching}

During the exploratory analysis process, there is relatively high probability that the user will also want to explore input one-step higher or lower in the structural hierarchy (i.e., whole residue for an observed atom and vice versa).
For example, the analyst may inspect a residue and observe an unexpected side chain movement, which he or she subsequently wants to investigate in more detail. 
For this reason, we introduce the predictive caching approach based on the structure hierarchy, caching not only the selected structure, but also parent structures (i.e., atom, residue, and ultimately chain).
Having the required information cached on different levels of detail allows users to freely traverse the data hierarchy without any significant latency.
We applied the principles of singleton caching for the overlapping input structures, as is often the case with large selections or whole chains accessed by multiple \smolboxes{}.

\subsection{Approximate RMSD Computation}
\label{sec:approximation}

The main purpose of \inputSmolbox{}es is to represent the input for subsequent \functionSmolbox{}es.
In addition, \inputSmolbox{}es also contribute to the analysis by providing information about the temporal development of selected parts of the molecule. 
This aproximate RMSD is computed as a central position for the input structure and indicates the center's deviation from its initial position in each snapshot.
This trade-off between speed and accuracy of displayed representation should serve solely as an instant indication of current trends in the selected input data.
Precise RMSD can be computed using dedicated \functionSmolbox{}.

\subsection{Pre-Computation and Custom CSV Files}
\label{sec:precomputation}

Despite our efforts, some measures are too complex to be computed on-the-fly, e.g., the detection of molecular tunnels and their properties, which may require hours of computation on a supercomputer.
Therefore, we allow users to load CSV (Comma Separated Values) files containing time series generated by other dedicated applications
or custom scripts. 
The CSV loader is designed in a simple manner: in the first line of the file it expects a list of measures to be loaded, then reads each remaining line as a snapshot index and the corresponding values for the measures.
This way, our tool can be easily extended and used for various analysis tasks for which it was not designed initially.

\subsection{Implementation}
\label{sec:implementation}

Implementation-wise, the application is written as a module in the CAVER Analyst visualization software \cite{jurcik2018caver}, using the Java programming language.
Therefore, it acts as a standalone, multiplatform application, mainly targeted to the use on personal computers.
The rendering engine and optimised cache is developed in-house, with focus on performance and memory limitations of personal computers.
The case studies, described in the following section, were performed on MacBook Pro with M1 ARM architecture and 16 GB of RAM.

\section{Case Studies}
\label{sec:case_study}

We performed two case studies with a senior protein engineer (10+ years of experience) using typical datasets from his daily work.
Over a long-term collaboration, this domain expert was involved in discussions about the protein engineering workflow, but he was not involved in the \smolboxes{} design prior to participating in the case studies. He is a co-author of the paper---a common collaboration approach for design studies~\cite{sedlmair2012design}.
For the first case study, the protein engineer revisited a previously published research based on two simulation ensembles~\cite{kaushik2018impact}.
In this study, he and his colleagues analyzed the binding of three molecules in the substrate 1-chlorohexane (CHA). They were initially placed in the solvent, to the haloalkane dehalogenase enzyme variant DhaA31, and the release of the respective dehalogenation products 1-hexanol (HAOL) and chloride ion (Cl\textsuperscript{--}) from the active site to the solvent.
These processes were compared with the dynamics of the protein tunnel (i.e., opening and closing of its entrance and bottleneck).
For the second case study, the protein engineer studied an unpublished unfolding process of two proteins---the wild type and its more stable mutant variant---to characterize their differences.

For the first case study, we used two simulations ensembles, each consisting of four 200 ns simulations with one snapshot per 2 ps (i.e., 400,000 snapshots in total). 
For the second case study, multiple parallel short simulations were run in adaptive rounds, for a total combined simulation time of 20 $\mu$s (200,000 snapshots spaced by 0.1 ns).

The focus of the case studies was twofold: 1) visual confirmation of the previous findings to validate the appropriateness of \smolboxes{} for exploratory MD simulation analysis and 2) a follow-up qualitative analysis for in-depth explanations of observed events and potential hypotheses for follow-up research. The data sets were chosen because they are relatively small and therefore enable a fluid analysis even on a commodity laptop.
At the same time, the simulations are sufficiently long so that watching the 3D MD simulation visualization is no longer feasible with any of the currently available tools, and scrubbing through the animation is far too imprecise.
In addition, the investigation focus is very different for the two cases---analyzing the interaction between structures vs.~comparing conformational changes over time---to demonstrate the versatility of the \smolboxes{} approach.

The studies lasted between one and two hours each.
The protein engineer was asked to prepare a summary of the already performed analyses, as well as a script describing the previously performed analysis steps to be replicated using \smolbox{} for the validation part of the study.
He was encouraged to think aloud while performing the investigation.
We performed screen- and audio-recording and transcribed the audio track. 
The transcript was coded into five categories: 1) questions / hypotheses about the data, 2) explanation of the (traditional) workflow, 3) findings and insights, 4) positive aspects about the interface mentioned, and 5) problems with the interface or suggestions for improvement.

\begin{figure}[tb]
  \centering
  \includegraphics[width=1.0\linewidth]{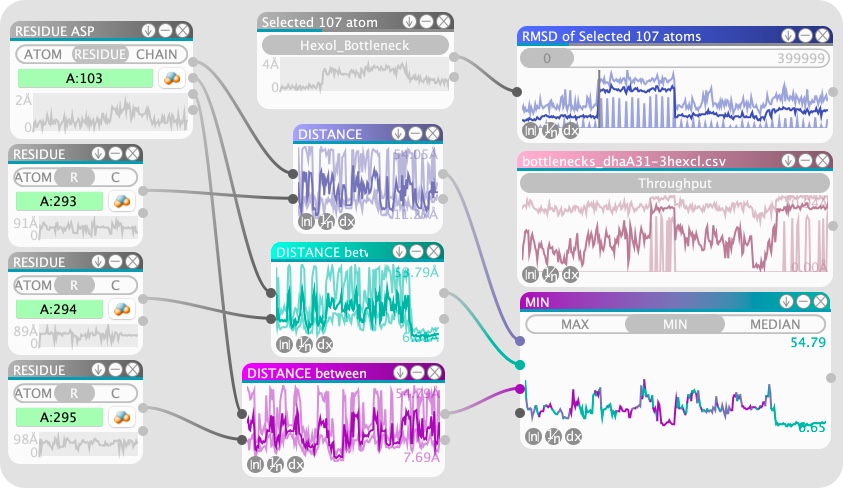}
  \caption{A setup of the network during the first study. The rightmost sMolBoxes show from the top: the RMSD of the residues forming the tunnel bottleneck, throughput of the tunnel bottleneck loaded from an external CSV file, and comparison of the minimal distance of three ligands (IDs 293, 294, 295) to the residue in the active site (ID 103).}
    \label{fig:CS_distances}       
\end{figure}

At the beginning of both case studies, the protein engineer tried to replicate previous analyses for  \textbf{visual confirmation}. In the first case study, the first step was to measure the distances between any of the three ligands to a a catalytic residue of the active site. Figure~\ref{fig:CS_distances} shows the three distance \functionSmolbox{}es, which were then combined using a minimum \aggregationSmolbox{} (bottom right). This way, he could quickly spot snapshots where at least one ligand was close to the active site.
He particularly appreciated the color-coding of the aggregated function, which allowed him to immediately see which ligand was currently the closest one to the active site.
The second confirmation step was to investigate if there is a correlation between the tunnel bottleneck and the ligands' distance to the active site. This would indicate that the presence of the ligand influenced the tunnel radius, which would be the expected behavior. 
By juxtaposing the minimum distance with the bottleneck radius, he could quickly confirm this expected correlation, which indicates that the presence of the ligand influences the tunnel radius (Figure~\ref{fig:CS_distances}).
These analyses were very similar to their traditional workflow. However, the protein engineer found it more intuitive, mainly because he could easily exchange the input residues by directly selecting a structure in a potentially interesting region identified in the 3D view.

In the second case study, the first step of the traditional analysis pipeline has been the computation of a Markov state model to construct different conformational states based on the RMSD of the entire protein structures.
Computing the RMSD for hundreds of thousands of snapshots for such large structures would violate the real-time analytics requirement of \smolboxes{} (R5).
Therefore, the protein engineer instead chose to compute distances from the start and end residues of the protein helices to a stable central residue of the protein.
The helices and the central residue are known for this protein.
Distance computations can be conducted on-the-fly, and the protein engineer confirmed that the shape of the curves resembles the RMSD curves: initially, the distances are low, and they increase over time.

\begin{figure}[tb]
  \centering
  \includegraphics[width=1.0\linewidth]{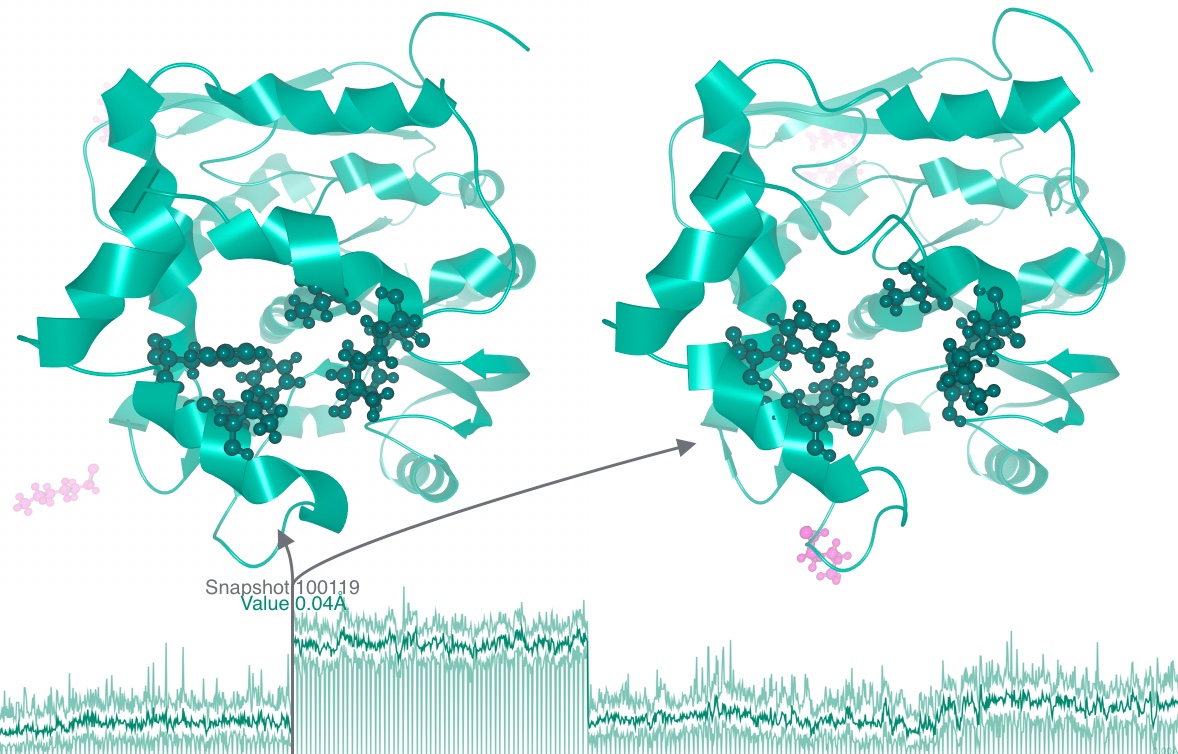}
  \caption{Sudden change in the tunnel bottleneck residue conformation near the snapshot 100,000. The top left shows the 3D View of the protein, highlighting those bottleneck residues (dark green) in a closed conformation. The top right shows the identical residues in the open conformation and the ligand (violet) approaching the tunnel entrance. The corresponding RMSD measure of the bottleneck residues is situated at the bottom.}
  \vspace{-3pt}
    \label{fig:CS_change}       
\end{figure}

For both case studies, the \textbf{qualitative follow-up investigation} yielded new insights. In the first case study, the protein engineer investigated the influence of the ligands on the bottleneck residues.
The bottleneck residues have been identified in prior analyses~\cite{kaushik2018impact} and are therefore known for this protein.
For the first two investigated bottleneck residues, he could quickly see a significant conformational change in longer time scale sequences of the simulations in the RMSD approximation curve.
This was unexpected but highly relevant since conformational changes of bottleneck residues can impact the binding of molecules and the opening/closing of the tunnel.
He confirmed this observation with a more precise RMSD function box (Figure~\ref{fig:CS_distances} top right).
By navigating to the snapshots where the conformation changed, he could observe a flipping of the two investigated bottleneck residues (Figure~\ref{fig:CS_change}).
He could confirm that one of the ligands was close to the tunnel mouth when the conformation change was registered.
That indicates that the presence of the ligand induces the bottleneck re-orientation.
For a more thorough follow-up analysis, he would investigate if multiple bottleneck residues open the tunnel in a concerted manner (synchronized) or if they move independently.

For the second case study, the protein engineer wanted to explore which helices initiate the unfolding process, which is not well understood yet.
By juxtaposing the pair-wise distance computations between all selected helix residues and the central residue, he concluded that some residues of the cap domain probably initiate the unfolding.
Other helices followed the unfolding later, but fluctuate even more in the end.
Compared to Markov state models, the protein engineer found it easier to understand the time-wise mechanism behind the unfolding by observing the distances over time and short simulation sequences associated with peak distances in the 3D View.
Their traditional Markov state analysis only delivers the states, but the temporal sequence gets lost.

\section{Discussion \& Future Work}
Overall, the protein engineer performing the case studies was intrigued by the potential of \smolboxes{}.
He particularly appreciated the rich options to perform on-the-fly analyses, such as quick comparisons between a variety of functions (R2) for any spatial selections (R1).
He noted, for instance, that he would not have considered performing a systematic RMSD analysis of bottleneck or catalytic residues before. 
However, the approximated RMSD preview of \inputSmolbox{}es made him aware of several properties worth further investigation, such as an unexpected instability of the active site, which is unusual since active site residues normally do not shift very much.
These discoveries were only made due to the instant computations and visualization within \smolboxes{} (R5, R6). 
\vspace{2mm}

To our surprise, our case study participant did not strictly follow the dataflow model from input to \aggregationSmolbox{} (R3) but rather opportunistically jumped between \smolboxes{} for setting different investigation foci and the linked views (R7).
Our prototype supported this behavior well. It illustrates the importance of having the option to quickly shift the focus forth and back---which would not be so easily possible in more linear analysis approaches, such as in Python notebooks.

We also did not expect the strong desire to analyze and compare the line charts directly within the Canvas and the importance of their spatial alignment. The protein engineer suggested providing more powerful methods to assess correlations---either through an easier superimposition of multiple function boxes or by a correlation aggregation that would compute the correlation of sliding time windows between two functions.

From the case studies, we could also observe that the analyst sometimes wished to perform a broader analysis comprising a larger number of structures, such as finding bottleneck, catalytic, or helix residues behaving in a similar way to hypothesize about entire functional units.
As \smolboxes{} operate on individual selections, such larger-scale comparisons are currently not supported.
From a conceptual point of view, we could enable such a broader analysis by performing batch operations on every substructure of a selection, for instance computing the RMSD on all residues of a selected protein chain.

Furthermore, the dataflow interface currently provides information about data provenance (R4). 
However, in the future, we would also like to extend the system with user action provenance to provide information about any modifications of the dataflow during the analysis.
We aim to use the provenance information combined with automated data analysis and event detection to train a recommendation system that could suggest potentially interesting data inputs, properties, and aggregations and guide the experts through the exploration process. We also plan to enable users to define their own sMolBoxes by combining multiple existing boxes into a tree that could be "collapsed" and stored as a single sMolBox for later use.

\section{Conclusion}

In this paper, we have presented \smolboxes{}, a novel system for visual exploration of long sequences of molecular dynamics (MD) simulations.
The proposed approach supports the workflow of computational chemists, enzymologists, and protein engineers and integrates quantitative analysis of molecular dynamics simulations into the visual exploratory analysis.
Our work takes inspiration from dataflow diagrams and enables fast selection of molecular inputs and computation, aggregation, and visualization of their associated properties.
Instead of obtaining measures for pre-defined structures through scripting, \smolboxes{} provide a powerful and flexible method for the discovery and qualitative characterization of important aspects of the MD simulations.

Two case studies with a protein engineer indicate that \smolboxes{} is flexible and responsive enough to support rich on-the-fly exploratory analysis to generate new hypotheses.
In the studies, \smolboxes{} enabled a fluid interplay between targeted MD simulation analyses using quantitative analysis with function plots and qualitative observations in a 3D visualization, even for simulations consisting of hundreds of thousands of snapshots.

\acknowledgments{
The authors wish to thank Sergej Stoppel, Robin Sk\r{a}nberg, and Mathieu Linares for their participation in the initial phases of the project.
The authors would also like to express their gratitude to the Czech Ministry of Education (INBIO - CZ.02.1.01/0.0/0.0/16\_026/0008451;
ELIXIR - LM2018131;
eINFRA - LM2018140)
and the Czech Science Foundation (20-15915Y).}

\bibliographystyle{abbrv-doi}

\bibliography{template}

\begin{thebibliography}{10}

\bibitem{Alharbi2016}
N.~Alharbi, R.~S. Laramee, and M.~Chavent.
\newblock {MolPathFinder: Interactive Multi-Dimensional Path Filtering of
  Molecular Dynamics Simulation Data}.
\newblock In C.~Turkay and T.~R. Wan, eds., {\em Computer Graphics and Visual
  Computing (CGVC)}. The Eurographics Association, 2016.

\bibitem{bidmon2008}
K.~Bidmon, S.~Grottel, F.~B\"{o}s, J.~Pleiss, and T.~Ertl.
\newblock Visual abstractions of solvent pathlines near protein cavities.
\newblock {\em Computer Graphics Forum}, 27(3):935--942, 2008.

\bibitem{brezovsky2018computational}
J.~Brezovsky, B.~Kozlikova, and J.~Damborsky.
\newblock Computational analysis of protein tunnels and channels.
\newblock In {\em Protein Engineering}, pp. 25--42. Springer, 2018.

\bibitem{bryden2012}
A.~Bryden, G.~Phillips, and M.~Gleicher.
\newblock Automated illustration of molecular flexibility.
\newblock {\em IEEE Transactions on Visualization and Computer Graphics},
  18(1):132--145, 2012.

\bibitem{byvska2019analysis}
J.~By{\v{s}}ka, T.~Trautner, S.~M. Marques, J.~Damborsk{\'y},
  B.~Kozl{\'\i}kov{\'a}, and M.~Waldner.
\newblock Analysis of long molecular dynamics simulations using interactive
  focus+ context visualization.
\newblock In {\em Computer Graphics Forum}, vol.~38, pp. 441--453. Wiley Online
  Library, 2019.

\bibitem{dabney2004mastering}
J.~B. Dabney and T.~L. Harman.
\newblock {\em Mastering simulink}, vol. 230.
\newblock Pearson/Prentice Hall Upper Saddle River, 2004.

\bibitem{duran2019}
D.~Duran, P.~Hermosilla, T.~Ropinski, B.~Kozl\'{i}kov\'{a}, A.~Vinacua, and
  P.-P. V\'{a}zquez.
\newblock Visualization of large molecular trajectories.
\newblock {\em IEEE Transactions on Visualization and Computer Graphics},
  25(1):987--996, 2019.

\bibitem{elmqvist2008datameadow}
N.~Elmqvist, J.~Stasko, and P.~Tsigas.
\newblock Datameadow: a visual canvas for analysis of large-scale multivariate
  data.
\newblock {\em Information Visualization}, 7(1):18--33, 2008.

\bibitem{Ertl2014}
T.~Ertl, M.~Krone, S.~Kesselheim, K.~Scharnowski, G.~Reina, and C.~Holm.
\newblock Visual analysis for space-time aggregation of biomolecular
  simulations.
\newblock {\em Faraday Discussions}, 01 2014.

\bibitem{Furmanova2017}
K.~Furmanov{\'a}, M.~Jare{\v{s}}ov{\'a}, J.~By{\v{s}}ka, A.~Jur{\v{c}}{\'i}k,
  J.~Parulek, H.~Hauser, and B.~Kozl{\'i}kov{\'a}.
\newblock Interactive exploration of ligand transportation through protein
  tunnels.
\newblock {\em BMC Bioinformatics}, 18 Suppl 2, 2017.

\bibitem{Hensen2012}
U.~Hensen, T.~Meyer, J.~Haas, R.~Rex, G.~Vriend, and H.~Grubm{\"u}ller.
\newblock {{E}xploring protein dynamics space: the dynasome as the missing link
  between protein structure and function}.
\newblock {\em PLoS ONE}, 7(5):e33931, 2012.

\bibitem{Hollingsworth2018}
S.~A. Hollingsworth and R.~O. Dror.
\newblock Molecular dynamics simulation for all.
\newblock {\em Neuron}, 99(6):1129--1143, 2018.

\bibitem{javed2013explates}
W.~Javed and N.~Elmqvist.
\newblock Explates: spatializing interactive analysis to scaffold visual
  exploration.
\newblock In {\em Computer Graphics Forum}, vol.~32, pp. 441--450. Wiley Online
  Library, 2013.

\bibitem{inviwo}
D.~J{\"o}nsson, P.~Steneteg, E.~Sund{\'e}n, R.~Englund, S.~Kottravel, M.~Falk,
  A.~Ynnerman, I.~Hotz, and T.~Ropinski.
\newblock Inviwo-a visualization system with usage abstraction levels.
\newblock {\em IEEE Transactions on Visualization and Computer Graphics}, 2019.

\bibitem{jurcik2018caver}
A.~Jur\v{c}\'{i}k, D.~Bedn\'{a}\v{r}, J.~By\v{s}ka, S.~M. Marques,
  K.~Furmanov\'{a}, L.~Daniel, P.~Kokkonen, J.~Brezovsk\'{y}, O.~Strnad,
  J.~\v{S}toura\v{c}, et~al.
\newblock Caver {Analyst} 2.0: analysis and visualization of channels and
  tunnels in protein structures and molecular dynamics trajectories.
\newblock {\em Bioinformatics}, 34(20):3586--3588, 2018.

\bibitem{kaushik2018impact}
S.~Kaushik, S.~M. Marques, P.~Khirsariya, K.~Paruch, L.~Libichov\'{a},
  J.~Brezovsk\'{y}, Z.~Prokop, R.~Chaloupkov\'{a}, and J.~Damborsk\'{y}.
\newblock Impact of the access tunnel engineering on catalysis is strictly
  ligand-specific.
\newblock {\em The FEBS journal}, 285(8):1456--1476, 2018.

\bibitem{kodosky2020labview}
J.~Kodosky.
\newblock Labview.
\newblock {\em Proceedings of the ACM on Programming Languages}, 4(HOPL):1--54,
  2020.

\bibitem{kokkonen2018molecular}
P.~Kokkonen, J.~S\'{y}kora, Z.~Prokop, A.~Ghose, D.~Bedn\'{a}\v{r}, M.~Amaro,
  K.~Beerens, v.~Bidmanov\'{a}, M.~Sl\'{a}nsk\'{a}, J.~Brezovsk\'{y},
  J.~Damborsk\'{y}, and M.~Hof.
\newblock Molecular gating of an engineered enzyme captured in real time.
\newblock {\em Journal of the American Chemical Society}, 140(51):17999--18008,
  2018.

\bibitem{koudelakova2013}
T.~Koudel\'{a}kov\'{a}, R.~Chaloupkov\'{a}, J.~Brezovsk\'{y}, Z.~Prokop,
  E.~\v{S}ebestov\'{a}, M.~Hesseler, M.~Khabiri, M.~Plevaka, D.~Kulik,
  I.~Kut\'{a}~Smatanova, P.~\v{R}ez\'{a}\v{c}ov\'{a}, R.~Ettrich, U.~T.
  Bornscheuer, and J.~Damborsk\'{y}.
\newblock Engineering enzyme stability and resistance to an organic cosolvent
  by modification of residues in the access tunnel.
\newblock {\em Angewandte Chemie International Edition}, 52(7):1959--1963,
  2013.

\bibitem{marques2017enzyme}
S.~M. Marques, L.~Daniel, T.~Buryska, Z.~Prokop, J.~Brezovsk\'{y}, and
  J.~Damborsk\'{y}.
\newblock Enzyme tunnels and gates as relevant targets in drug design.
\newblock {\em Medicinal Research Reviews}, 37(5):1095--1139, 2017.

\bibitem{marques2017catalytic}
S.~M. Marques, Z.~Dunajov\'{a}, Z.~Prokop, R.~Chaloupkov\'{a},
  J.~Brezovsk\'{y}, and J.~Damborsk\'{y}.
\newblock Catalytic cycle of haloalkane dehalogenases toward unnatural
  substrates explored by computational modeling.
\newblock {\em Journal of Chemical Information and Modeling}, 57(8):1970--1989,
  2017.

\bibitem{martinez2020visualizing}
X.~Martinez, M.~Chavent, and M.~Baaden.
\newblock Visualizing protein structures—tools and trends.
\newblock {\em Biochemical Society Transactions}, 48(2):499--506, 2020.

\bibitem{mei2018design}
H.~Mei, Y.~Ma, Y.~Wei, and W.~Chen.
\newblock The design space of construction tools for information visualization:
  A survey.
\newblock {\em Journal of Visual Languages \& Computing}, 44:120--132, 2018.

\bibitem{spradow2009}
J.~{Meyer-Spradow}, T.~{Ropinski}, J.~{Mensmann}, and K.~{Hinrichs}.
\newblock Voreen: A rapid-prototyping environment for ray-casting-based volume
  visualizations.
\newblock {\em IEEE Computer Graphics and Applications}, 29(6):6--13, 2009.

\bibitem{patro2011}
R.~Patro, C.~Y. Ip, S.~Bista, D.~Thirumalai, S.~S. Cho, and A.~Varshney.
\newblock {MDMap}: A system for data-driven layout and exploration of molecular
  dynamics simulations.
\newblock In {\em 2011 IEEE Symposium on Biological Data Visualization
  (BioVis).}, pp. 111--118, 2011.

\bibitem{patro2010}
R.~Patro, C.~Y. Ip, A.~Varshney, and H.~Hagen.
\newblock Saliency guided summarization of molecular dynamics simulations.
\newblock {\em Scientific Visualization: Advanced Concepts}, 1:321--335, 2010.

\bibitem{pavlova2009}
M.~Pavlov\'{a}, M.~Klva\v{n}a, R.~Chaloupkov\'{a}, P.~Ban\'{a}\v{s},
  M.~Otyepka, R.~Wade, Y.~Nagata, and J.~Damborsk\'{y}.
\newblock Redesigning dehalogenase access tunnels as a strategy for degrading
  an anthropogenic substrate.
\newblock {\em Nature Chemical Biology}, (5):727--733, 2009.

\bibitem{mevislab}
F.~Ritter, T.~Boskamp, A.~Homeyer, H.~Laue, M.~Schwier, F.~Link, and H.-O.
  Peitgen.
\newblock Medical image analysis.
\newblock {\em IEEE Pulse}, 2(6):60--70, 2011.

\bibitem{schatz2021visual}
K.~Schatz, J.~J. Franco-Moreno, M.~Sch{\"a}fer, A.~S. Rose, V.~Ferrario,
  J.~Pleiss, P.-P. V{\'a}zquez, T.~Ertl, and M.~Krone.
\newblock Visual analysis of large-scale protein-ligand interaction data.
\newblock In {\em Computer Graphics Forum}, vol.~40, pp. 394--408. Wiley Online
  Library, 2021.

\bibitem{sedlmair2012design}
M.~Sedlmair, M.~Meyer, and T.~Munzner.
\newblock Design study methodology: Reflections from the trenches and the
  stacks.
\newblock {\em IEEE transactions on visualization and computer graphics},
  18(12):2431--2440, 2012.

\bibitem{Silva2007}
C.~T. {Silva}, J.~{Freire}, and S.~P. {Callahan}.
\newblock Provenance for visualizations: Reproducibility and beyond.
\newblock {\em Computing in Science Engineering}, 9(5):82--89, 2007.

\bibitem{Skanberg2018}
R.~Sk\r{a}nberg, M.~Linares, C.~K\"{o}nig, P.~Norman, D.~J\"{o}nsson, I.~Hotz,
  and A.~Ynnerman.
\newblock {VIA-MD: Visual Interactive Analysis of Molecular Dynamics}.
\newblock In J.~Byska, M.~Krone, and B.~Sommer, eds., {\em Workshop on
  Molecular Graphics and Visual Analysis of Molecular Data}. The Eurographics
  Association, 2018. doi: {{%
10\hspace{.1pt}\discretionary{.}{%
}{.}\hspace{.4pt}2312\discretionary{/}{%
}{/}molva\hspace{.1pt}\discretionary{.}{%
}{.}\hspace{.4pt}20181102}}


\bibitem{vvvv}
J.~{Someone} and T.~{Something}.
\newblock {VVVV} a multipurpose toolkit.
\newblock \url{https://vvvv.org/}, 2019.
\newblock Accessed: 2019-03-09.

\bibitem{Upson1989}
C.~{Upson}, T.~A. {Faulhaber}, D.~{Kamins}, D.~{Laidlaw}, D.~{Schlegel},
  J.~{Vroom}, R.~{Gurwitz}, and A.~{van Dam}.
\newblock The application visualization system: a computational environment for
  scientific visualization.
\newblock {\em IEEE Computer Graphics and Applications}, 9(4):30--42, 1989.

\bibitem{vad2017watergate}
V.~Vad, J.~By{\v{s}}ka, A.~Jur\v{c}{\'\i}k, I.~Viola, E.~Gr{\"o}ller,
  H.~Hauser, S.~M. Marques, J.~Damborsk{\`y}, and B.~Kozl{\'\i}kov{\'a}.
\newblock Watergate: Visual exploration of water trajectories in protein
  dynamics.
\newblock In {\em Proceedings of Eurographics Workshop on Visual Computing for
  Biology and Medicine (EG VCBM)}, pp. 33--42, 2017.

\bibitem{vazquez2018}
P.-P. V\'{a}zquez, P.~Hermosilla, V.~Guallar, J.~Estrada, and A.~Vinacua.
\newblock Visual analysis of protein-ligand interactions.
\newblock {\em Computer Graphics Forum}, 37(3):391--402, 2018.

\bibitem{waser2011nodes}
J.~Waser, H.~Ribicic, R.~Fuchs, C.~Hirsch, B.~Schindler, G.~Bloschl, and
  E.~Gr\"{o}ller.
\newblock Nodes on ropes: A comprehensive data and control flow for steering
  ensemble simulations.
\newblock {\em IEEE Transactions on Visualization and Computer Graphics},
  17(12):1872--1881, 2011.

\bibitem{Yu2017}
B.~{Yu} and C.~T. {Silva}.
\newblock Visflow - web-based visualization framework for tabular data with a
  subset flow model.
\newblock {\em IEEE Transactions on Visualization and Computer Graphics},
  23(1):251--260, 2017.

\bibitem{yu2019flowsense}
B.~Yu and C.~T. Silva.
\newblock Flowsense: A natural language interface for visual data exploration
  within a dataflow system.
\newblock {\em IEEE Transactions on Visualization and Computer Graphics},
  26(1):1--11, 2019.

\end{thebibliography}
\end{document}